\documentclass{article}
\usepackage{xcolor}
\usepackage[margin=1in]{geometry}
\usepackage{amsmath,verbatim}
\usepackage{amssymb}
\usepackage{graphicx}
\usepackage{subcaption}
\usepackage{float}
\usepackage{amsthm}
\usepackage{array} 
\newcommand{\D}{\mathrm{D}}
\newcommand{\N}{\mathrm{N}}
\newcommand{\A}{\mathrm{A}}
\newcommand{\B}{\mathrm{B}}

\newcommand{\ri}{\mathrm{ri}}
\newcommand{\ro}{\mathrm{ro}}

\newtheorem{theorem}{Theorem}
\newtheorem{remark}{Remark}

\usepackage{graphicx} 
\usepackage{amsfonts, amsmath, amssymb, amsthm}
\usepackage{url}

\title{A theory of ROC analysis of  rule-out and rule-in diagnostics with applications to mammography data}
\author{Michelle Mastrianni$^{1}$, Kwok Lung Fan$^{1}$, Yee Lam Elim Thompson$^{1}$, \\ Jessie J.J. Gommers$^{2}$, Ioannis Sechopoulos$^{2}$, Fredrik Strand$^{3}$, Weijie Chen$^{1}$, \\ Gary Levine$^{1}$, Mukul Sherekar$^{1}$, Frank W. Samuelson$^{1*}$ \\
\\
$^{1}$ The U.S. Food and Drug Administration\\
$^{2}$ Department of Medical Imaging, Radboud University Medical Center, \\ Nijmegen, the Netherlands\\
$^{3}$ Department of Oncology-Pathology, Karolinska Institute, Stockholm, Sweden\\
$^{*}$ Correspondence: frank.samuelson@fda.hhs.gov}
\date{\today}

\begin{document}
\maketitle



\begin{abstract}
Multiple diagnostic tests are frequently used to determine the presence
of a disease condition in patients.  In this paper, we use bivariate
copulas to examine the properties of receiver operating characteristic (ROC) curves formed when two
correlated diagnostic tests are used together to rule-out (``believe the
negative'') and rule-in (``believe the positive'') patients for disease.  We
use this theory to analyze three mammography data sets
where AI devices are applied to reduce radiologists' workload or
improve diagnostic performance.  Our analysis shows with generality
that increasing the radiologist-AI correlation for diseased cases
enhances the area under the ROC curve (AUC) of a radiologist-AI rule-out curve, whereas decreasing correlation
for non-diseased cases has a similar effect. The opposite trends hold
for rule-in scenarios. Applications to clinical mammography data show
that projected empirical radiologist performance under a rule-out or
rule-in scenario is consistent with the theory.
\end{abstract}

\section{Introduction}
\label{sec:introduction}

Multiple diagnostic tests are frequently used to determine the presence~($D$)
or absence~($N$) of disease conditions in patients.  Many authors have devised methods to evaluate multiple diagnostics, and typically these methods assume
both tests yield binary outputs ~\cite{Marshall, Obuchowski, McClish}.
Frequently binary tests are used in one of two combinations when the tests disagree. In a ``rule-in'' or ``believe the positive'' scenario (BP), patients are considered positive if either of the tests is positive. In a ``rule-out'' or ``believe the negative'' scenario (BN), patients are considered negative if either of the tests is negative. Various authors have compared statistical metrics for BN and BP tests with a control arm~\cite{Khreich, Mahkonen, McClish, Shen, Thompson, Marshall}.

Diagnostic tests with ordinal or quantitative outputs may be evaluated using receiver operating characteristic (ROC) curves, which represent all possible true and false positive rates of the test as the decision threshold is varied.  
In the area of radiological imaging, ROC curves are often used to evaluate radiologist performance~\cite{Metz2007} because (1) humans can discern different levels of disease likelihood and (2) human decision thresholds vary from person to person, with disease prevalence, and with perceived utilities such as financial costs or benefit to the patient ~\cite{Metz2006}.

Given two tests $A$ and $B$ with ordinal or quantitative outputs, a typical method to evaluate receiver operator characteristic (ROC) curves in the BN or BP setting is to evaluate the MROC (maximum ROC) curve.  MROC chooses the pair of score thresholds for $A$ and $B$ which maximize the sensitivity of the joint distribution (based either on BN or BP) for every specificity ~\cite{Khreich, Shen, Thompson}.  However, such maximization of a ROC curve is not possible with many
kinds of diagnostic tests such as commercially purchased artificial intelligence (AI) algorithms or human radiologists.

In this paper we describe properties of joint ROC curves where two tests have ordinal or quantitative outputs, and Test~A is used at 
a fixed threshold to
rule-in or rule-out cases for evaluation by a second test~B.  The forms of these curves depend upon the
performance of the tests and the correlation between tests.  
Of interest to us is the scenario where Test~A is a radiological AI device, and
Test~B is a radiologist whose evaluations are correlated with the AI device dependent on disease
status ($D$ or $N$).

The field of medical imaging has seen transformative advancements with the use of artificial intelligence (AI) algorithms to improve the efficacy and efficiency of disease screening. In the United States, the standard of care for medical imaging interpretation typically involves a reading by a single radiologist. Thus, FDA-authorized AI devices are primarily intended to be used as a diagnostic aid or a tool to triage image review priority. Various studies have illuminated the benefits of AI imaging devices in mammography~\cite{Larsen, McKinney, RodriguezRuiz, Santeramo, vanWinkel}, lung cancer screening~\cite{Burns, Qin}, and other modalities.

One option to reduce radiologist workload is to automatically ``rule-out'' AI-negative patients as non-diseased, therefore bypassing the need for a radiologist review. A number of retrospective studies have shown the promise of such devices in reducing workload while maintaining or improving other metrics of performance in mammography screening \cite{Hickman, Kyono, RayaPovedano, RodriguezRuiz2}. In many European countries, the standard-of-care includes a double-reading by two radiologists, typically with the expectation of reaching a consensus on diagnostic conclusion. Due to this rigorous standard of care, conducting prospective studies on rule-out devices is more easily facilitated in these countries. Several such studies are in progress or recently completed~\cite{Lang}.

With single-radiologist standard-of-care, an AI-negative patient may not be read by human readers with the use of a rule-out device. Therefore, evaluation metrics for
future rule-out devices may differ from those used for existing devices, which are currently intended mainly as a second reader to aid or triage patients. Obuchowski \cite{Obuchowski} compares various performance metrics, including sensitivity, specificity, negative predictive value (NPV), and negative likelihood ratio (NLR), in the rule-out setting and the radiologist-alone setting, assuming no correlation between the radiologist and AI readers. We aim to expand on these statistical analyses by considering the effect of the diseased and non-diseased correlations between the radiologist and AI diagnostics.



In \S \ref{sec:theory_intro}, we give a brief introduction to ROC analysis and present several different types of bivariate copulas for constructing a joint distribution from arbitrary marginals. In \S \ref{sec:theory_correlation} we expand on the theory for rule-out, rule-in, and combined scenarios, and evaluate the effect of the diseased and non-diseased correlations on the area under the ROC curve (AUC) of the joint distribution under the different scenarios. After developing the theory, we turn in \S \ref{sec:empirical} to an analysis of several real-world mammography datasets with radiologist screening scores and AI scores. We show that our copula models agree well with empirical rule-out and rule-in data for these datasets and compare several metrics between the radiologist-alone ROC curve and the joint radiologist-AI diagnostic.

\section{Univariate ROC theory and bivariate copulas}
\label{sec:theory_intro}

This section will introduce the necessary fundamentals for modeling rule-out and rule-in ROC curves. We start with the univariate ROC theory and then give a brief introduction to copulas, including several common copulas which we shall use in our models.

\subsection{ROC curves from distributions}

A common practice in ROC analysis is to presume that the test results can be transformed to fit some  distributional form and to construct a smooth ROC curve based on the assumed model~\cite{Metz}. A typical assumption is that the test scores are monotonically transformable to two binormal distributions, though in some fields another distribution may be a more suitable fit. Typically, the score distribution for non-diseased patients is assumed without loss of generality to be centered at 0, while the diseased patients have a higher mean score.

\begin{figure}[h]
\begin{center}
\includegraphics[scale=0.7]{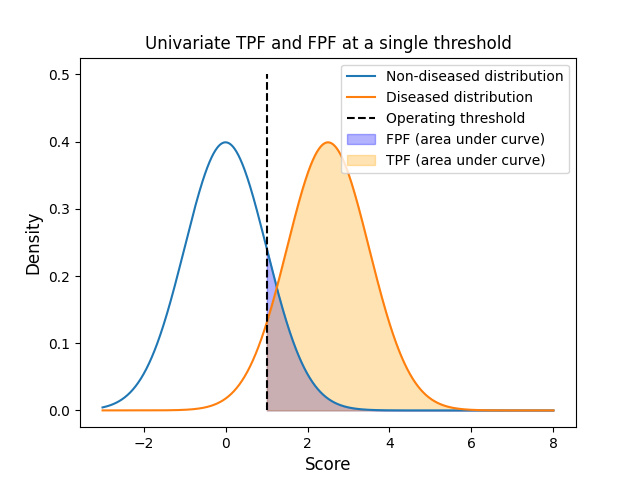}
\end{center}
\caption{Use of a bi-normal distribution to obtain the false positive fraction (FPF) and true positive fraction (TPF) at a particular threshold}
\label{fig:1}
\end{figure}

Suppose that the test outputs of the non-diseased and diseased patients have underlying distributions with cumulative distribution functions (CDFs) $F_{\N}$ and $F_{\D}$ respectively. The true positive fraction (TPF) of an ROC curve is plotted against the false positive fraction (FPF). In this case, we have for a threshold $t$:
\begin{align*} 
&\mathrm{FPF}(t) = 1-F_{\N}(t), \text{ and}\\
&\mathrm{TPF}(t) = 1-F_{\D}(t).
\end{align*}
In other words, the FPF and TPF at threshold $t$ are determined by the survival function (the complement to the CDF) of the non-diseased and diseased distributions respectively at that threshold. See Figure \ref{fig:1} for a graphical depiction where the score distributions are binormal.

\subsection{Joint distributions via bivariate copulas}
\label{subsec:theory_intro_copulas}
\begin{figure}[H]
\begin{center}
\includegraphics[scale=0.6]{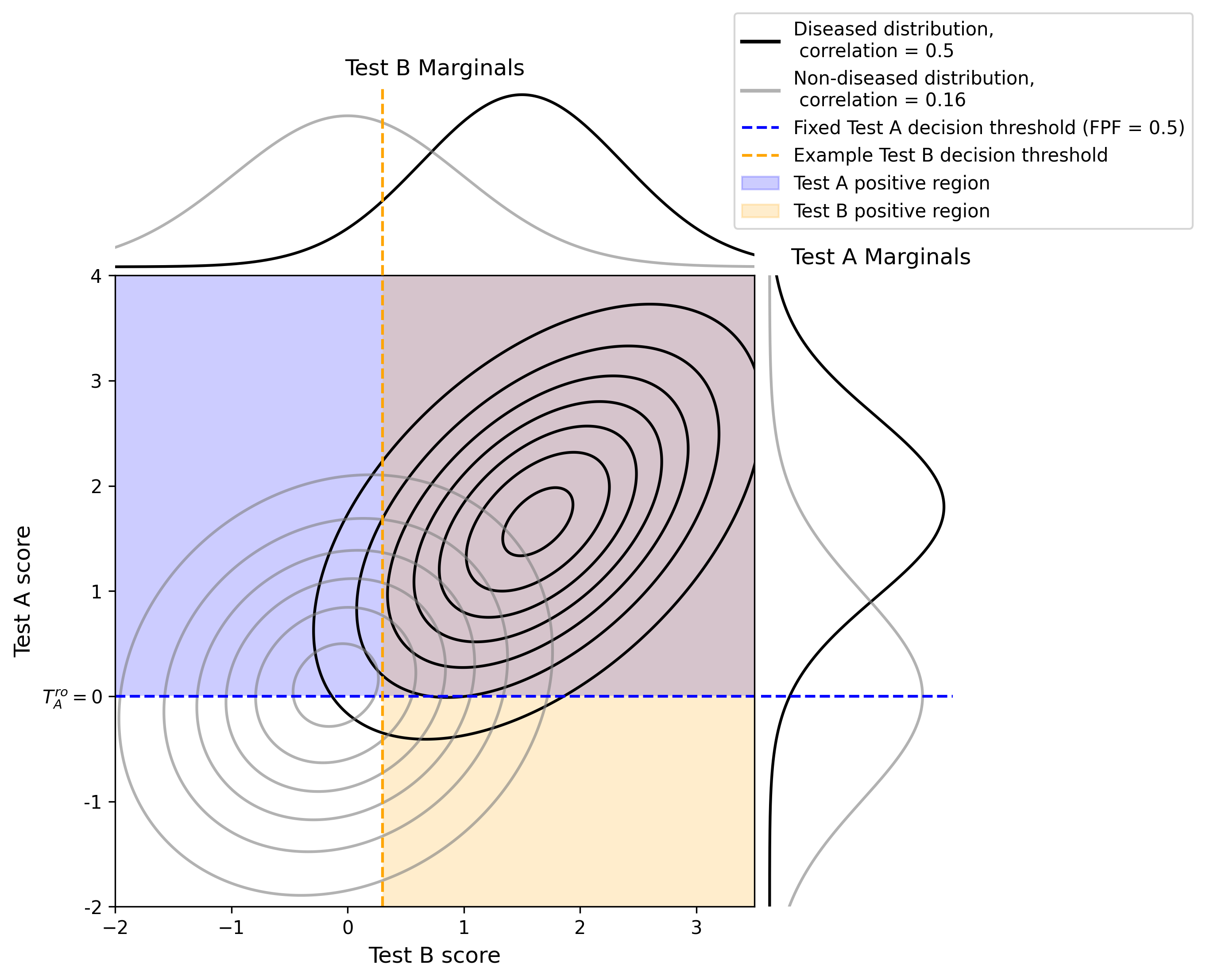}
\end{center}
\caption{An example of Gaussian copula density plots for a joint diagnostic setting. Contours representing the population density of test scores from non-diseased patients are shown in gray and the diseased distribution is shown in black. The test marginal distributions are binormal. Pearson's correlations between the tests on non-diseased patients is 0.16 and on diseased patients is 0.5. For visualization purposes, example Test~A and Test~B decision thresholds are set by the blue and orange dashed lines respectively. This model is very similar to that used in CORROC~\cite{CORROC}.}
\label{fig:joint_copulas} 
\end{figure}

In \S \ref{sec:theory_correlation}, we will use \emph{bivariate copulas} to construct joint  distributions for two tests with given marginals. The general definition of a copula~\cite{Takeuchi} is as follows.  Let $G$ be a joint distribution function with margins $F_A$ and $F_B$. By Sklar's theorem, there is a bivariate copula $C$ such that for all $x, y \in \mathbb{R}$
$$G(x,y) = C[F_A(x),F_B(y)] := C(u, v),$$
where $u = F_A(x)$ and $v = F_B(y)$. Figure \ref{fig:joint_copulas} illustrates joint (rule-out) score distributions from Test~A (AI) and Test~B (radiologist) and marginals using a Gaussian copula structure as detailed below in \S\ref{subsec:gaussian}.  Non-diseased distribution contours are shown in gray and diseased distributions in black. The upper region (blue shaded) is considered positive by Test~A.  The right region (yellow shaded) is considered positive by Test~B.
In a rule-in/BP scenario, cases from all shaded areas will be considered positive.  In a rule-out/BN scenario only cases in the region with brown (yellow+blue) shading are positive.
False positives are reduced significantly in a rule-out scenario when compared to either the Test-B-alone or Test-A-alone scenarios.

Below, we detail a few specific well-known copulas that we will use for our rule-out and rule-in ROC models.

\subsubsection{Gaussian copulas}
\label{subsec:gaussian}
A Gaussian copula is constructed by transforming the variables to normal variables with mean zero and variance 1 (i.e., $\mathcal{N}(0,1)$) and then combining them under a bivariate normal CDF. Let $\psi_1$ and $\Psi_1$ respectively be the PDF and CDF of $\mathcal{N}(0,1)$: $$\psi_1(x) = \frac{1}{\sqrt{2\pi}} \exp\Big(\frac{-x^2}{2}\Big), \hspace{0.5cm} \Psi_1(x) = \int_{\infty}^x \psi_1(x') dx'.$$
Further let $\psi_2$ and $\Psi_2$ be the joint PDF and CDF
\begin{align*}\psi_2(x,y;\rho) & = \frac{1}{\sqrt{(2\pi)^2(1-\rho^2)}}\exp\Big[-\frac{x^2+y^2-2\rho x y}{2(1-\rho^2)}\Big], \\ \hspace{0.5cm} \Psi_2(x,y;\rho) &= \int_{-\infty}^{x} \int_{\infty}^{y} \psi_2(x',y';\rho) dx' dy',
\end{align*}
where the correlation $\rho$ is the Pearson's correlation coefficient.

Then the \emph{Gaussian copula} is defined as 
$$C_{\mathrm{Gaussian}}(u, v, \rho) = \Psi_2[\Psi_1^{-1}(u),\Psi_1^{-1}(v);\rho].$$
where~$u$ and~$v$ are the marginal probabilities, i.e. the values of the marginal CDFs. 

The Gaussian copula is flexible in terms of capturing linear dependence through the correlation matrix. Pearson’s rho is well-suited to the Gaussian copula because both rely on the assumption of linearity: the Pearson's rho is defined in terms of covariances and standard deviations, which are inherently linear concepts, and this measure of correlation quantifies the degree to which two variables change together linearly.

\subsubsection{Archimedean copulas}
Archimedean copulas encompass a family of copulas with shared properties. Several such copulas (for example, the Gumbel, Clayton, and Frank copulas) are well-studied in the literature~\cite{Naifar}. Many common Archimedean copulas have an explicit formula. This gives an advantage over the Gaussian copula, as there is no way to express the CDF of a Gaussian in closed form. Further, while the copula function of the Gaussian distribution assumes tail independence---meaning that extreme events in one variable are not associated with extreme events in the other---some Archimedean copulas can indeed capture dependence in the upper or lower tails. 

An Archimedean copula with generator function $\phi$ can be written as follows:
$$C_{\mathrm{arch}}(u, v) = \phi^{-1}[\phi(u) + \phi(v)].$$
Here, $\phi(z)$ is a continuous, strictly decreasing and convex function such that $\phi(1) = 0$ and $\phi(0) = \infty$, and $\phi^{-1}$ has the same properties (except that $\phi^{-1}(\infty) = 0$ and $\phi^{-1}(0) = 1$).

Some common Archimedean copulas and  their generator functions are given in the table below. We will analyze the effect of the correlation on the ROC curves generated by these copulas in the next section.

\begin{table*}[h]
    \centering
    \renewcommand{\arraystretch}{1.7} 
    \begin{tabular}{|c|c|c|} 
        \hline
         & Copula & Generator $\phi(z)$ \\
        \hline
        \textbf{Gumbel} & $C_{\mathrm{Gumbel}} = \exp\big[-\big((-\log u)^{\theta}+(-\log v)^{\theta}\big)^{1/\theta} \big]$ & $(-\log z)^{\theta}$  \\
        \hline
        \textbf{Clayton} & $C_{\mathrm{Clayton}} = \big[\max(u^{-\theta}+v^{-\theta}-1,0)\big]^{-1/\theta}$ & $z^{-\theta}-1$ \\
        \hline
        \textbf{Frank} & $C_{\mathrm{Frank}} = -\frac{1}{\theta}\log\big[1+\frac{(e^{-\theta u}-1)(e^{-\theta v}-1)}{e^{-\theta}-1}\big]$ &  $-\log\big(\frac{e^{-\theta z}-1}{e^{-\theta}-1}\big)$\\
        \hline
    \end{tabular}
    \caption{Types of Archimedean Copulas}
    \label{tab:copulas} 
\end{table*}

Since the generator function $\phi$ typically depends on a parameter $\theta$, we will use the notation $C(u, v; \theta)$ where $\theta$ is a parameter related to the correlation,similar to $\rho$ in the Gaussian case.





\section{Rule-out, rule-in, and combination ROC curves from copulas}
\label{sec:theory_correlation}

We now examine a joint Test~A/B 
diagnostic in both rule-out and rule-in scenarios, as well as settings that integrate both rule-out and rule-in scenarios simultaneously.
In the rule-out scenario, we adopt a ``believe the negative'' approach under which A-negative cases 
are not passed to Test~B. 
Thus, in a rule-out setting, a case is deemed negative if either Test~A or~B is 
negative. In the rule-in scenario, images deemed positive by the Test~A are automatically called positive (without the need for Test~B), in a ``believe the positive" approach.  In a combined scenario, some percentage of high-confidence A-negative patients are ruled out from follow-up, while a percentage of high-confidence A-positive patients are automatically called back. See, for example,~\cite{Hickman} for analysis of such a combined scenario in mammography imaging.

We will generate the joint distributions via copulas of the various types introduced in \S\ref{subsec:theory_intro_copulas} and analyze the impact of the 
correlations between Tests~A and~B on the resulting ROC curves formed from those distributions. We introduce the following notation which will be used through the rest of the section.

\begin{itemize}
    \item The Test-B univariate marginal CDF functions are denoted $F_{B\N}(x)$ and $F_{B\D}(x)$ respectively for the non-diseased and diseased classes.
    \item Similarly, the Test-A CDF marginals are denoted $F_{A\N}(y)$ and $F_{A\D}(y)$.
    \item The clinical Test-A threshold settings are denoted as $T_{\A}^{\ro}$ and $T_{\A}^{\ri}$ for rule-out and rule-in respectively. Cases with Test-A score below $T_{\A}^{\ro}$ are ruled-out, while cases above $T_{\A}^{\ri}$ are ruled-in.
\end{itemize}

\begin{figure}[h]
\begin{center}
\includegraphics[scale=0.55]{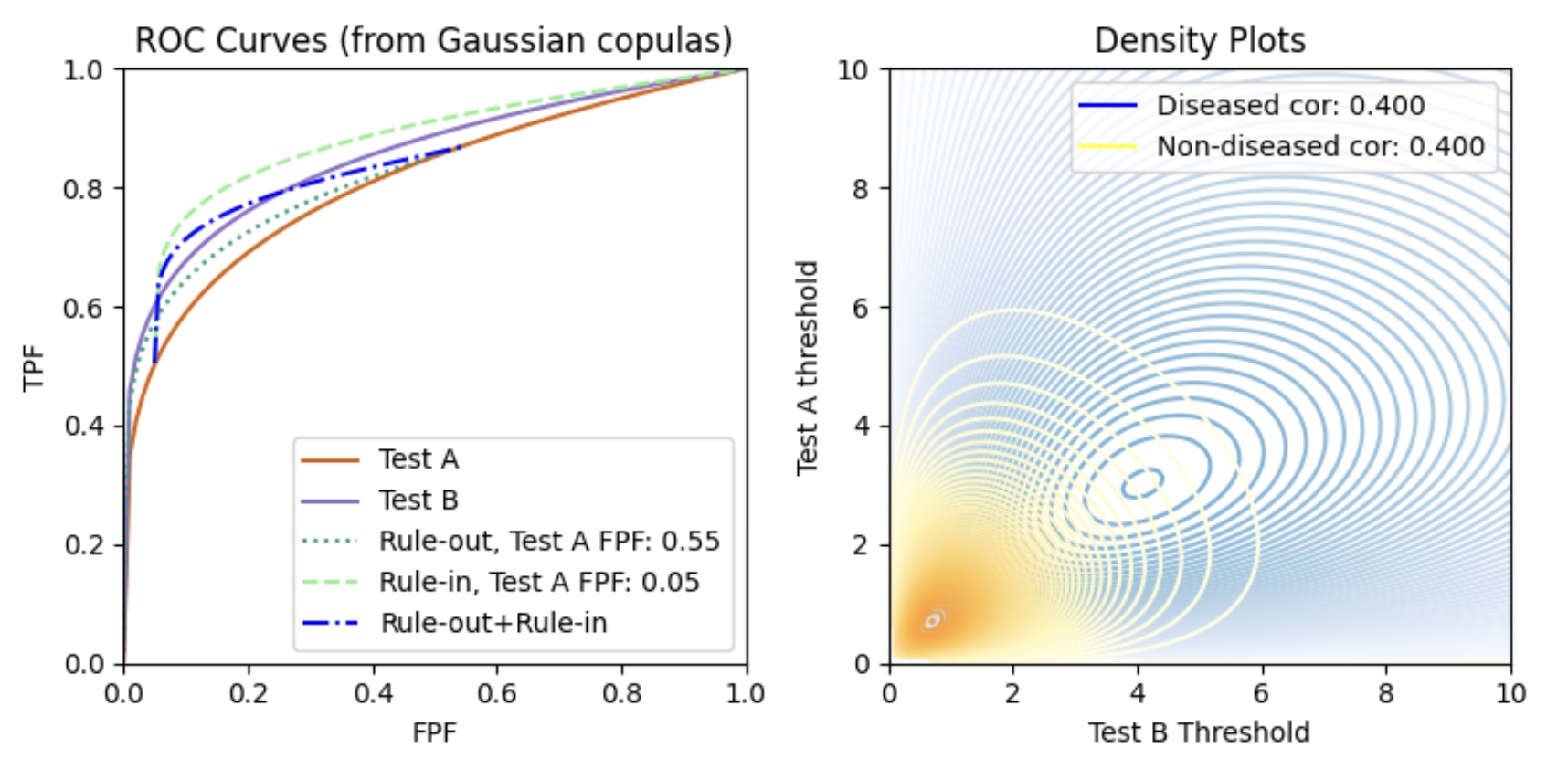}
\caption{On the left, we model the Test-A-alone (e.g. AI-alone), Test-B-alone (e.g. radiologist-alone),
rule-out, rule-in, and combination rule-out+rule-in ROC curves, constructed via a Gaussian copulas. We assume exponentially distributed marginals for diseased cases, $\mathrm{Exp}(\lambda)$, 
with parameters $\lambda = 0.23$ for Test~A and $\lambda = 0.17$ for Test~B. 
For non-diseased cases, we assume exponential distributions with $\lambda = 1$ for both tests.
The Pearson correlation coefficients are $\rho_{N} = 0.4$ for non-diseased cases and $\rho_{D} = 0.4$ for diseased cases.
The rule-out Test-A false-positive fraction (FPF) is set at $0.55$, and the rule-in FPF is $0.05$.  Note that the rule-out curve stops at $\text{FPF} = 0.55$, while the rule-in curve starts at $\text{FPF} = 0.05$.
The combination curve spans the FPF range [0.05, 0.55]. On the right, we show density plots for the transformed test scores in the latent Gaussian copula space, where each exponential score $x$ is mapped to a uniform variable via its corresponding exponential CDF $u=1-e^{-\lambda x}$ (with $\lambda$ depending on disease status and test) and then transformed as $z = \Phi^{-1}(u)$.}
\label{fig:ROC_ruleout_rulein}
\end{center}
\end{figure}

In contrast to  the maximum ROC curve \cite{McClish, Shen}, which is common in the literature for constructing ROC curves for joint diagnostics, our rule-out and rule-in ROC curves 
are constructed under the assumption that Test A 
operates at rule-out threshold $T_{\A}^{\ro}$ and rule-in threshold  $T_{\A}^{\ri}$. This results in an improper joint ROC curve whose FPF range is restricted: the largest possible FPF of the joint ROC curve is the FPF of the marginal A-alone ROC corresponding to threshold $T_{\A}^{\ro}$, which we refer to as $\mathrm{FPF_A}(T_{\A}^{\ro})=1-F_{\A\N}\left(T_{\A}^{\ro}\right)$, while $\mathrm{FPF_A}(T_{\A}^{\ri})$ is the smallest possible FPF in the joint rule-in-out scenario. See Figure \ref{fig:ROC_ruleout_rulein} for reference.
Note that a more complete representation of a two-test joint  diagnostic without a fixed AI operating point would live in three-dimensional ROC space~\cite{Shen}.

We are interested in how correlations affect the ROC curve generated from rule-in and rule-out scenarios. The  correlation measure used in our analyses will depend on the type of copula.
An important characteristic of multivariable distributions is the correlation measure between variables. Perhaps the most well-known are Pearson's product-moment correlation coefficient, which measures a linear dependence between two variables; Spearman's $\rho$, which is a non-parametric version of the linear correlation; and Kendall's $\tau$, which quantifies the similarity of the ordering of observations between two variables. These metrics capture different aspects of dependence (linear vs. rank-based). We are interested in how these correlation measures affect the ROC curve generated from a pair of copula-generated joint distributions. 

The correlation measure used in our analyses will depend on the type of copula. For simplicity, we use positive correlation measures (Pearson's rho or Kendall's tau), given that in our context a radiologist and AI are likely to have positive correlation.

\subsection{Rule-out scenario}

For each of the copula types discussed in \S \ref{subsec:theory_intro_copulas}, we will give precise expressions for the FPF and TPF for the rule-out ROC curves constructed from those copulas, and analyze the effect of changes in the diseased and non-diseased correlations on the curves. In particular, we are interested the effect of the correlations on the area under the curve (AUC), a common metric with higher values indicating more favorable diagnostic performance. In the rule-out and rule-in scenarios, we focus on the partial AUC (pAUC) over the restricted range of FPF values $[\mathrm{FPF_A}(T_{\A}^{\ro}), \mathrm{FPF_A}(T_{\A}^{\ri})]$ as described in \S \ref{sec:theory_correlation}.
$$\text{pAUC}=\int_{\mathrm{FPF_A}(T_{\A}^{\ro})}^{\mathrm{FPF_A}(T_{\A}^{\ri})} \text{TPF}\cdot \text{d FPF}.$$
Our main result is summarized in the following theorem.

\begin{theorem}
\label{thm:1}[Correlation effects on rule-out ROC curves]
For two diagnostic tests $A$ and $B$, define the Pearson's rho correlation of scores given by $A$ and $B$ on \emph{signal-present} or \emph{diseased} cases as $\rho_{\D}$ and the on \emph{signal-absent} or \emph{non-diseased} cases as $\rho_{\N}$. Similarly define the Kendall's tau correlations as $\tau_{\D}$ and $\tau_{\N}$, and the Spearman's rho correlations as $\rho_{S, \D}$, $\rho_{S, \N}$. Then,
\begin{enumerate}
    \item The pAUC of a rule-out ROC curve generated by a Gaussian copula increases with increasing $\rho_{\D}$ and with decreasing $\rho_{\N}$.
    \item The pAUC of a rule-out ROC curve generated by a Gumbel, Clayton, or Frank copula increases with increasing $\tau_{\D}$ and with decreasing $\tau_{\N}$.
    \item The pAUC of a rule-out ROC curve generated by any copula $C(u, v, \theta)$ for which $C$ is monotonic in the $\theta$ parameter increases with increasing $\rho_{S, \D}$ and with decreasing $\rho_{S, \N}$. 
\end{enumerate}
\end{theorem}
These effects of correlation on AUC were previously noted by Shen~\cite{Shen}.
The following subsections are dedicated to proving Theorem~\ref{thm:1}.

\subsubsection{Gaussian copulas}
In this section we prove part 1 of Theorem \ref{thm:1}.
We assume for simplicity that $\rho_{\N} > 0$ and $\rho_{\D} > 0$.

Because the FPF depends on the non-diseased marginals, we can express the FPF of the rule-out diagnostic as a function of the Test-B threshold at a fixed Test-A threshold of $T_{\A}^{ro}$ using the survival function of the Gaussian copula:
\begin{align}
\mathrm{FPF}&\vert_{\mathrm{Gaussian}}^{\mathrm{rule-out}}(x;y=T_{\A}^{\ro};\rho_{\N}) = \nonumber \\ 
&\int_{\Psi_1^{-1}(F_{B\N}(x))}^{\infty} \int_{\Psi_1^{-1}(F_{A\N}(T_{\A}^{ro}))}^{\infty} \frac{1}{\sqrt{2\pi^2(1-\rho_{\N})^2}} \exp\Big(-\frac{x'^2 + y'^2 - 2\rho_{\N}x'y'}{2(1-\rho_{\N})^2}\Big) dx' dy'.
\label{eq:FPFruleout_gauss}
\end{align}
Similarly,
\begin{align}
\mathrm{TPF}&\vert_{\mathrm{Gaussian}}^{\mathrm{rule-out}}(x;y=T_{\A}^{\ro};\rho_{\D}) = \nonumber\\ 
&\int_{\Psi_1^{-1}(F_{\B\D}(x))}^{\infty} \int_{\Psi_1^{-1}(F_{\A\D}(T_{\A}))}^{\infty} \frac{1}{\sqrt{2\pi^2(1-\rho_{\D})^2}} \exp\Big(-\frac{x'^2 + y'^2 - 2\rho_{\D}x'y'}{2(1-\rho_{\D})^2}\Big) dx' dy'.
\label{eq:TPFruleout_gauss}
\end{align}

To see how FPF depends on $\rho_{\N}$ from \eqref{eq:FPFruleout_gauss}, we introduce an identity~\cite{DW,GF} to obtain another expression for $\mathrm{FPF}\vert_{\mathrm{Gaussian}}^{\mathrm{rule-out}}(x;y=T_{\A}^{\ro};\rho_{\N})$:
\begin{align}
\mathrm{FPF}\vert_{\mathrm{Gaussian}}^{\mathrm{rule-out}}&(x;y=T_{\A}^{\ro};\rho_{\N}) = \Psi_1(-F_{\B\N}(x))\Psi_1(-F_{\A\N}(T_{\A}^{\ro})) \nonumber \\
& +\frac{1}{2\pi} \int_{0}^{\rho_{\N}} \frac{1}{\sqrt{1-t^2}} \exp\Big(\frac{-F_{\B\N}(x)^2-2F_{\B\N}(x)F_{\A\N}(T_{\A}^{\ro})+F_{\A\N}(T_{\A}^{\ro})^2}{2(1-t)^2}\Big) dt.
\label{eq:FPFruleout_gauss_id2}
\end{align}
And again, similar for the TPF we can write
\begin{align}
\mathrm{TPF}\vert_{\mathrm{Gaussian}}^{\mathrm{rule-out}}&(x;y=T_{\A}^{\ro};\rho_{\D}) = \Psi_1(-F_{\B\D}(x))\Psi_1(-F_{\A\D}(T_{\A}^{\ro})) \nonumber\\
& +\frac{1}{2\pi} \int_{0}^{\rho_{\D}} \frac{1}{\sqrt{1-t^2}} \exp\Big(\frac{-F_{\B\D}(x)^2-2F_{\B\D}(x)F_{\A\D}(T_{\A}^{\ro})+F_{\A\D}(T_{\A}^{\ro})^2}{2(1-t)^2}\Big) dt.
\label{eq:TPFruleout_gauss_id2}
\end{align}

Seeing the effect of an increase in the diseased correlation $\rho_{\D}$ on the joint ROC curve is relatively simple from \eqref{eq:TPFruleout_gauss_id2}. At a fixed Test-B  score $x$, increasing $\rho_{\mathrm{D}}$ simply increases the upper bound of the integral in \eqref{eq:TPFruleout_gauss_id2} and since the integrand is positive, the TPF increases. Thus, for $\rho_{\D,1} > \rho_{\D,2}$
$$\mathrm{TPF}\vert_{y=T_{\A}^{\ro}}^{\mathrm{rule-out}}(x;\rho_{\D,1}) > \mathrm{TPF}\vert_{y=T_{\A}^{\ro}}^{\mathrm{rule-out}}(x;\rho_{\D,2}),$$
while the FPF is fixed (assuming no changes in $\rho_{\N}$). Hence, the sensitivity of the joint rule-out distribution increases at each value of the specificity with an increase in $\rho_{\D}$, and the pAUC increases.

An increase in the non-diseased correlation, $\rho_{\N}$, has the opposite effect. 
For a fixed TPF, the FPF of the joint rule-out distribution increases with an increase in $\rho_{\N}$, leading to a \emph{decrease} in the pAUC.

\subsubsection{Archimedean copulas}
Here we prove part 2 of Theorem \ref{thm:1}. Unlike the Gaussian copula, whose distributions are expressed in terms of Pearson's $\rho$, it is not as easy to see an explicit dependence on correlation for Archimedean copulas based on the generator functions. However, there is an explicit relationship between Kendall's $\tau$ and the generating function $\phi$ due to Genest and MacKay \cite{GM}:
\begin{align}
    \tau = 4\int_0^1 \frac{\phi(z)}{\phi'(z)} dz +1
    \label{eq:tautheta_relationship}
\end{align}

For the \textbf{Gumbel copula}, the integrand in   \eqref{eq:tautheta_relationship} is $\frac{\phi(z)}{\phi'(z)} = \frac{z \ln z}{\theta}$. We can write Kendall's $\tau$ as
\begin{align}
\label{eq:tau_gumbel}
    \tau_{\mathrm{Gumbel}}(\theta) = 4 \int_0^1 \frac{z \ln z}{\theta} dz +1 = 1-\frac{1}{\theta}.
\end{align}
When $\theta \in (1, \infty)$, we have $0 < \tau_{\mathrm{Gumbel}} < 1$, which is our region of interest. Equation \eqref{eq:tau_gumbel} is clearly increasing in $\theta$ on $(1, \infty)$. Thus, if $\tau(\theta_1) > \tau(\theta_2)$ with $\mathrm{sgn}(\theta_1) = \mathrm{sgn}(\theta_2)$, then $\theta_1 > \theta_2$, i.e. increasing $\theta$ increases correlation as 
measured by Kendall's $\tau$.

Now, suppose that we construct a joint Test-A-B 
ROC curve from marginals $F_{B\N}$, $F_{B\D}, F_{A\N}$, and $F_{A\D}$ using non-diseased and diseased copulas with parameters $\theta_{\N}$ and $\theta_{\D}$ respectively:
\begin{align}
&\mathrm{FPF}\vert_{\mathrm{Gumbel}}^{\mathrm{rule-out}}(x;y=T_{\A}^{\ro};\theta_{\N}) \nonumber\\
&=1-F_{\B\N}(x)-F_{\A\N}(T_{\A}^{\ro})+\exp\big[-\big((-\log F_{\B\N}(x))^{\theta_{\N}}+(-\log F_{\A\N}(T_{\A}^{\ro}))^{\theta_{\N}}\big)^{1/\theta_{\N}} \big]
\label{eq:FPFruleout_gumbel}
\end{align}
and
\begin{align}
\mathrm{TPF}&\vert_{\mathrm{Gumbel}}^{\mathrm{rule-out}}(x;y=T_{\A}^{\ro};\theta_{\D}) \nonumber\\
&=1-F_{\B\D}(x)-F_{\A\D}(T_{\A}^{\ro}) + \exp\big[-\big((-\log F_{\B\D}(x))^{\theta_{\D}}+(-\log F_{\A\D}(T_{\A}^{\ro}))^{\theta_{\D}}\big)^{1/\theta_{\D}} \big].
\label{eq:TPFruleout_gumbel}
\end{align}
Note that the above survival functions are based on the inclusion-exclusion principle for bivariate distributions~\cite{DW}. After some calculus (see \S\ref{subsec:A1}), it is clear that Equation~\ref{eq:TPFruleout_gumbel} increases in $\theta_{\D}$ on $(1, \infty)$. If $\tau(\theta_{\D 1}) > \tau(\theta_{\D 2})$, then $\theta_{\D 1} > \theta_{\D 2}$ and by the increasing property
$$\mathrm{TPF}\vert_{\mathrm{Gumbel}}^{\mathrm{rule-out}}(x_1;y=T_{\A}^{\ro};\theta_{\D,1}) > \mathrm{TPF}\vert_{\mathrm{Gumbel}}^{\mathrm{rule-out}}(x_2;y=T_{\A}^{\ro};\theta_{\D,2}).$$
In other words, an increase in Kendall's tau, assuming no change in sign, leads to an increase in TPF for fixed FPF, and thus an increase in pAUC.

As with the Gaussian copula, for fixed sensitivity, an increase in the non-diseased Kendall's $\tau$ (that is, an increase in $\theta_{\N})$ leads to an increase in the FPF, and thus a decrease in pAUC.\\

For the \textbf{Clayton copula}, the integrand in     \eqref{eq:tautheta_relationship} is
$\frac{\phi(z)}{\phi'(z)} = \frac{z^{\theta+1}-z}{\theta}$, so
\begin{align}
\label{eq:tau_clayton}
    \tau_{\mathrm{Clayton}}(\theta) &= 4 \int_0^1 \frac{z^{\theta+1}-z}{\theta} dz +1\\
    & = \frac{\theta}{\theta+2} \text{ for } \theta > -2.
\end{align}
When $\theta \in (0, \infty)$, we have $0 < \tau_{\mathrm{Clayton}} < 1$, which is our region of interest. This function is increasing in $\theta$ for $\theta > 0$ and fixed $z \in (0,1)$. So as $\theta$ increases, Kendall's tau $\tau_{\mathrm{Clayton}}$ increases. The TPF is given by 
\begin{align}
\mathrm{TPF}&\vert_{\mathrm{Clayton}}^{\mathrm{rule-out}}(x;y=T_{\A}^{\ro};\theta_{\D}) \nonumber\\
&=1-F_{\B\D}(x)-F_{\A\D}(T_{\A}^{\ro}) + [\max(F_{\B\D}(x)^{-\theta_{\D}} + F_{\A\D}(T_{\A}^{\ro})^{-\theta_{\D}} -1, 0)]^{-1/\theta_{\D}}.
\label{eq:TPFruleout_clayton}
\end{align}
For $\theta_{\D} > 0$, $[\max(F_{\B\D}(x)^{-\theta_{\D}} + F_{\A\D}(T_{\A}^{\ro})^{-\theta_{\D}} -1, 0)] = F_{\B\D}(x)^{-\theta_{\D}} + F_{\A\D}(T_{\A}^{\ro})^{-\theta_{\D}}$. We can then deduce that \eqref{eq:TPFruleout_clayton} is increasing in $\theta_{\D}$ via calculus very similar to the arguments in \S\ref{subsec:A1}. By repeating essentially the same arguments as for the previous copulas, 
we deduce the same conclusion that an increase in Kendall's diseased $\tau_{\mathrm{Clayton}}$ leads to an increase in rule-out sensitivity for fixed specificity, while an increase in non-diseased Kendall's $\tau_{\mathrm{Clayton}}$ leads to a decrease in sensitivity.\\

Finally, to analyze the \textbf{Frank copula}, rather than using the relationship given in \eqref{eq:tautheta_relationship}, we use the identity
\begin{align}
\tau_{\mathrm{Frank}} = 1 + \frac{4}{\theta}\Big(D_1(\theta)-1\Big),
\label{eq:taufrank}
\end{align}
\cite{Naifar}, where $D_1(\theta)$ is the Debye function $\frac{1}{\theta} \int_0^{\theta} \frac{t}{e^t -1} dt$. When $\theta \in (0, \infty)$, we have $0 < \tau_{\mathrm{Frank}} < 1$, which is the region of interest. This theta-tau relationship can be explained using the Series expansion for the Debye function (see \S \ref{subsec:A2} which shows that $\tau_{\mathrm{Frank}}$ is increasing in $\theta$ for $\theta > 0$). The TPF is given by 
\begin{align}
\mathrm{TPF}&\vert_{\mathrm{Frank}}^{\mathrm{rule-out}}(x;y=T_{\A}^{\ro};\theta_{\D}) \nonumber\\
&=1-F_{\B\D}(x)-F_{\A\D}(T_{\A}^{\ro}) -\frac{1}{\theta_{\D}}\log\Big[1+\frac{(e^{-\theta_{\D} F_{\B\D}(x)}-1)(e^{-\theta_{\D} F_{\A\D}(T_{\A}^{\ro})}-1)}{e^{-\theta_{\D}}-1}\Big].
\label{eq:TPFruleout_frank}
\end{align}
This is increasing in $\theta_{\D}$ (see \S\ref{subsec:A3}), and again we have that an increase in diseased Kendall's tau leads to an increase in the pAUC, while an increase in non-diseased Kendall's tau leads to a decrease in the pAUC. 

\subsubsection{Spearman's rho}

Parts (i) and (ii) of Theorem \ref{thm:1} focus on specific copulas and their corresponding correlation measures. Here we prove part (iii), which gives the same conclusion when the correlation measure is Spearman's rho, for \emph{any} copula $C$ which is monotonic in a correlation parameter $\theta$. This thus generalizes the results from Parts  (i) and (ii) of Theorem \ref{thm:1}.

From Theorem 5.1.6 of Nelsen~\cite{Nelsen}, the following relationship holds for a general copula $C(u, v; \theta)$ and Spearman's rho $\rho_{S}(C)$:
\begin{align}
\label{eq:spearman_copula}
\rho_{S}(C) =  12 \int_0^1 \int_0^1 C(u, v; \theta) du dv - 3.
\end{align}
The function $C(u, v; \theta)$ is nonnegative everywhere. Thus, if $C(u, v; \theta)$ is monotonically increasing (or decreasing) in $\theta$, so is the right-hand side of \eqref{eq:spearman_copula}, and thus $\rho_{S}(C)$. Therefore,
\begin{align}
\label{eq:rho_C_relationship}
\frac{d C}{d \rho_S} = \frac{\frac{d C}{d\theta}}{\frac{d \rho_S}{d \theta}} > 0.
\end{align}

As in \eqref{eq:FPFruleout_gumbel} and \eqref{eq:TPFruleout_gumbel}, the bivariate FPF and TPFs at a Test-A threshold $T_{\A}$ are given respectively by
\begin{align}
\label{eq:FPF_general_copula}
&\mathrm{FPF}\vert_{C}^{\mathrm{rule-out}}(x;y=T_{\A}^{\ro};\theta_{\N}) \nonumber\\
&=1-F_{\B\N}(x)-F_{\A\N}(T_{\A}^{\ro})+C(F_{\B\N}(x), F_{\A\N}(T_{\A}^{\ro}); \theta_{\N}) 
\end{align}
and
\begin{align}
\label{eq:TPF_general_copula}
\mathrm{TPF}&\vert_{C}^{\mathrm{rule-out}}(x;y=T_{\A}^{\ro};\theta_{\D}) \nonumber\\
&=1-F_{\B\D}(x)-F_{\A\D}(T_{\A}^{\ro}) + C(F_{\B\D}(x), F_{\A\D}(T_{\A}^{\ro}); \theta_{\D})  
\end{align}

In \eqref{eq:FPF_general_copula}, the first three terms are not dependent on $\theta_{\N}$ and thus not dependent on the non-diseased Spearman's rho correlations $\rho_{S, \N}$. The same holds for \eqref{eq:TPF_general_copula} with $\theta_{\D}$ and the diseased Spearman's rho $\rho_{S, \D}$. The last terms in \eqref{eq:FPF_general_copula} and \eqref{eq:TPF_general_copula} are respectively increasing in $\rho_{S, \N}$ and $\rho_{S, \D}$ by \eqref{eq:rho_C_relationship}. Thus, as $\rho_{S, \N}$ increases, the FPF increases at a fixed TPF, leading to a decrease in the AUC; and as  $\rho_{S, \D}$ increases, the TPF increases at a fixed FPF, leading to an increase in the AUC.

\subsection{Rule-in scenario}

The rule-in scenario is straightforward to analyze in conjunction with rule-out. Recall that rule-in curves \emph{start} at the rule-in FPF threshold, as in Figure \ref{fig:ROC_ruleout_rulein}. As discussed, we may view rule-in as a \emph{believe the positive} scenario in which \emph{both} Test~A and Test~B 
must deem a patient positive in order to flag that patient positive. Note the contrast to rule-out, in which \emph{one of} the tests 
must deem a patient negative in order to flag that patient. We thus have by the inclusion-exclusion principle
\begin{align}
\label{eq:FPFrulein}
\mathrm{FPF}\vert^{\mathrm{rule-in}}(x;y=T_{\A};\rho_{\N}) &= (1-F_{\B\N}(x)) + (1-F_{\A\N}(T_{\A})) \nonumber \\
& \hspace{1cm} - \mathrm{FPF}^{\mathrm{rule-out}}(x;y=T_{\A};\rho_{\N}) \nonumber\\
\mathrm{TPF}\vert^{\mathrm{rule-in}}(x;y=T_{\A};\rho_{\D}) &= (1-F_{\B\D}(x)) + (1-F_{\A\D}(T_{\A})) \\
& \hspace{1cm} - \mathrm{TPF}\vert^{\mathrm{rule-out}}(x;y=T_{\A};\rho_{\D}).
\label{eq:FPFruleout}
\end{align}

The above relationship holds regardless of the copula structure for rule-out. From these equations, we can easily deduce that the correlations in diseased and non-diseased cases affect the rule-in ROC curve in opposite ways compared to the rule-out ROC curve: an increase in the diseased correlation results in a decrease in sensitivity at each specificity (thus decreasing the pAUC), whereas an increase in the non-diseased correlation leads to an increase in the pAUC. This is summarized in the following theorem.

\begin{theorem}
\label{thm:2}[Correlation effects on rule-in ROC curves]
For two diagnostic tests $A$ and $B$, define the Pearson's rho correlation of scores given by $A$ and $B$ on \emph{signal-present} or \emph{diseased} cases as $\rho_{\D}$ and the on \emph{signal-absent} or \emph{non-diseased} cases as $\rho_{\N}$. Similarly define the Kendall's tau correlations as $\tau_{\D}$ and $\tau_{\N}$, and the Spearman's rho correlations as $\rho_{S, \D}$, $\rho_{S, \N}$. Then,
\begin{enumerate}
    \item The pAUC of a rule-in ROC curve generated by a Gaussian copula increases with decreasing $\rho_{\D}$ and with increasing $\rho_{\N}$.
    \item The pAUC of a rule-in ROC curve generated by a Gumbel, Clayton, or Frank copula increases with decreasing $\tau_{\D}$ and with increasing $\tau_{\N}$.
    \item The pAUC of a rule-in ROC curve generated by any copula $C(u, v, \theta)$ for which $C$ is monotonic in the $\theta$ parameter increases with decreasing $\rho_{S, \D}$ and with increasing $\rho_{S, \N}$. 
\end{enumerate}
\end{theorem}

\subsection{Modeling combination rule-out and rule-in scenarios}

To model the joint ROC curve of a diagnostic that simultaneously employs both rule-out and rule-in methods as in Figure \ref{fig:ROC_ruleout_rulein}, we need the FPF
\begin{align} 
\label{eq:FPFriro}
\mathrm{FPF}_{\ri+\ro} = \frac{\mathrm{FP}_{\ri} + \mathrm{FP}_{\ro} - \mathrm{FP}_{\mathrm{int}}}{M_0},
\end{align}
where $\mathrm{FP}_{\ri}$ is the number of false positive patients from the rule-in diagnostic; $\mathrm{FP}_{\ro}$ is the number of false positive patients from the rule-out diagnostic; $\mathrm{FP}_{\mathrm{int}}$ is the number of false positive patients from the intersection of the two (in other words, double-counted false positives); and $M_0$ 
is the total number of non-diseased patients. 

As usual, the rule-in score threshold is $T_{\A}^{\ri}$, and the rule-out score threshold is $T_{\A}^{\ro} < T_{\A}^{\ri}$. We denote the Test-B threshold by $T_{\B}$. Then \eqref{eq:FPFriro} is simply
\begin{align} 
\mathrm{FPF}_{\ri+\ro}(x) & = \mathbb{P}(\text{Test-A score for ND image } > T_{\A}^{\ri}, \mathrm{\N}) \nonumber \\
& \hspace{1cm} + \mathbb{P}(\text{Test-A score for ND image } > T_{\A}^{\ro}, \text{Test-B score }x\text{ for ND image } > T_{\B}) \nonumber \\
& \hspace{1cm} - \mathbb{P}(\text{Test-A score for ND image } > T_{\A}^{\ri}, \text{Test-B score }x\text{ for ND image } > T_{\B}) \nonumber \\
& = (1-F_{\A\N}(T_{\A}^{\ri})) + \mathrm{FPF}^{\mathrm{rule-out}}(x;y=T_{\A}^{\ro};\rho_{\N}) - \mathrm{FPF}^{\mathrm{rule-in}}(x; y=T_{\A}^{\ri}; \rho_{\N}),
\label{eq:FPFriro2}
\end{align}
where as usual $F_{\A\N}$ is the marginal CDF of Test~A and the joint rule-out FPFs are given by some choice of copula. Similarly, 
\begin{align} 
\mathrm{TPF}_{\ri+\ro}(x) & = \mathbb{P}(\text{Test-A score for D image } > T_{\A}^{\ri}, \mathrm{\D}) \nonumber \\
& \hspace{1cm} + \mathbb{P}(\text{Test-A score for D image } > T_{\A}^{\ro}, \text{Test-B score }x\text{  for D image } > T_{\B}) \nonumber \\
& \hspace{1cm} - \mathbb{P}(\text{Test-A score for D image } > T_{\A}^{\ri}, \text{Test-B score }x\text{  for D image } > T_{\B}) \nonumber \\
& = (1-F_{\A \D}(T_{\A}^{ri})) + \mathrm{TPF}^{\mathrm{rule-out}}(x;y=T_{\A}^{\ro};\rho_{\D}) - \mathrm{TPF}^{\mathrm{rule-in}}(x; y=T_{\A}^{\ri}; \rho_{\D}).
\label{eq:FPFriro3}
\end{align}

Analyzing the impact of correlations in the combined scenario is more complex because the correlations influence the rule-out and rule-in scenarios in opposing ways.

\section{Validation of theory with empirical mammography datasets}
\label{sec:empirical}


To validate our theory, we analyzed several real-world mammography datasets to which AI algorithms (Test~A) have been applied and for which radiologist scores (Test~B) at screening were available. This allows us to evaluate radiologist performance assuming that the AI algorithm will be hypothetically used to rule-out and/or rule-in patients.  We can use any general diagnostic metric such as the AUC and/or sensitivity at a fixed specificity.

In each example, in order to construct estimated joint ROC curves from the empirical data, we need the AI-alone ROC curve, a radiologist ROC curve or an estimate of the curve using the radiologist TPF and FPF on the test set, and a measure of correlation between the AI and radiologists for both the non-diseased and diseased patients. 
To fit a smooth (parametric) ROC curve to an empirical ROC curve, we transformed the empirical ROC values using the inverse of the assumed cumulative distributions and then fit them with linear Deming regression. Our assumed cumulative distributions are binormal.

\begin{remark}
In each scenario considered, the efficacy of the device in terms of the percentage of patients ruled out from the worklist can be calculated as follows, where $p$ is the disease prevalence and the FPF and TPF are that of the AI device:
$$\% \text{ of patients ruled out} =(1-p) \cdot (1-\text{FPF})+p \cdot (1-\text{TPF}).$$
For a disease with very low prevalence in real-world scenarios, the \% of patients ruled out can thus be approximated by the specificity of the device.
\end{remark}


\subsection{Emory Breast Imaging Dataset}

The full EMory BrEast Imaging Dataset (EMBED) contains 3.4 million mammography studies from 110,000 patients collected from 2013-2020 at Emory University in Atlanta, Georgia, USA \cite{EMBED}. Images come from both screening and diagnostic exams. 
The open-source version of EMBED \cite{EMBED_opensource} contains roughly 10\% of the full dataset, which has 364,000 screening and diagnostic mammographic exams for 110,000 patients and pathology results.

\subsubsection{Dataset}

We filtered images in the open-source dataset as follows. First, we took the earliest screening date per patient (if it exists) as the ``initial exam." We discarded patients with only diagnostic exams, given that a patient typically bypasses a screening exam if presenting with symptoms. Ground truth for cancerous images was established via biopsy results, including the Breast Imaging Reporting and Data System (BIRADS) scores, within 180 days of the initial exam. Non-cancerous images were confirmed with negative screening results within 500 days of follow-up.

All patients for whom ground truth could not be established---for instance those who have no exams within 500 days of the initial screening exam---were excluded from the analysis. The final dataset consists of 91,809 images from 8,126 patients, 681 of which have confirmed cancerous pathology. 11,031 of the images were flagged abnormal at screening.

We randomly assigned roughly 50\% of the images to a training set, 25\% to a tuning set, and 25\% to a test set, keeping images corresponding to the same patient in the same set. We then randomly discarded non-cancerous images such that roughly 30\% of the images in each set were cancerous. Our final test set is comprised of 119 cancerous images and 271 non-cancerous images.

\subsubsection{AI algorithm}
The AI scores for the EMBED test set were obtained using the publicly available 2nd-place winning algorithm from the 2023 RSNA Screening Mammography Breast Cancer Detection competition hosted on Kaggle \cite{RSNAKaggle}. This algorithm uses PyTorch to enhance model performance through a series of retraining and fine-tuning steps. A single-view model is established and then refined to develop a dual-view model. The dual-view model also incorporates meta-information such as age, implant status, and machine ID to further refine detection capabilities.
 
\subsubsection{Analysis}

\begin{figure}[H]
\begin{center}
\includegraphics[trim=0cm 3cm 0cm 1cm, clip, width=14.5cm]{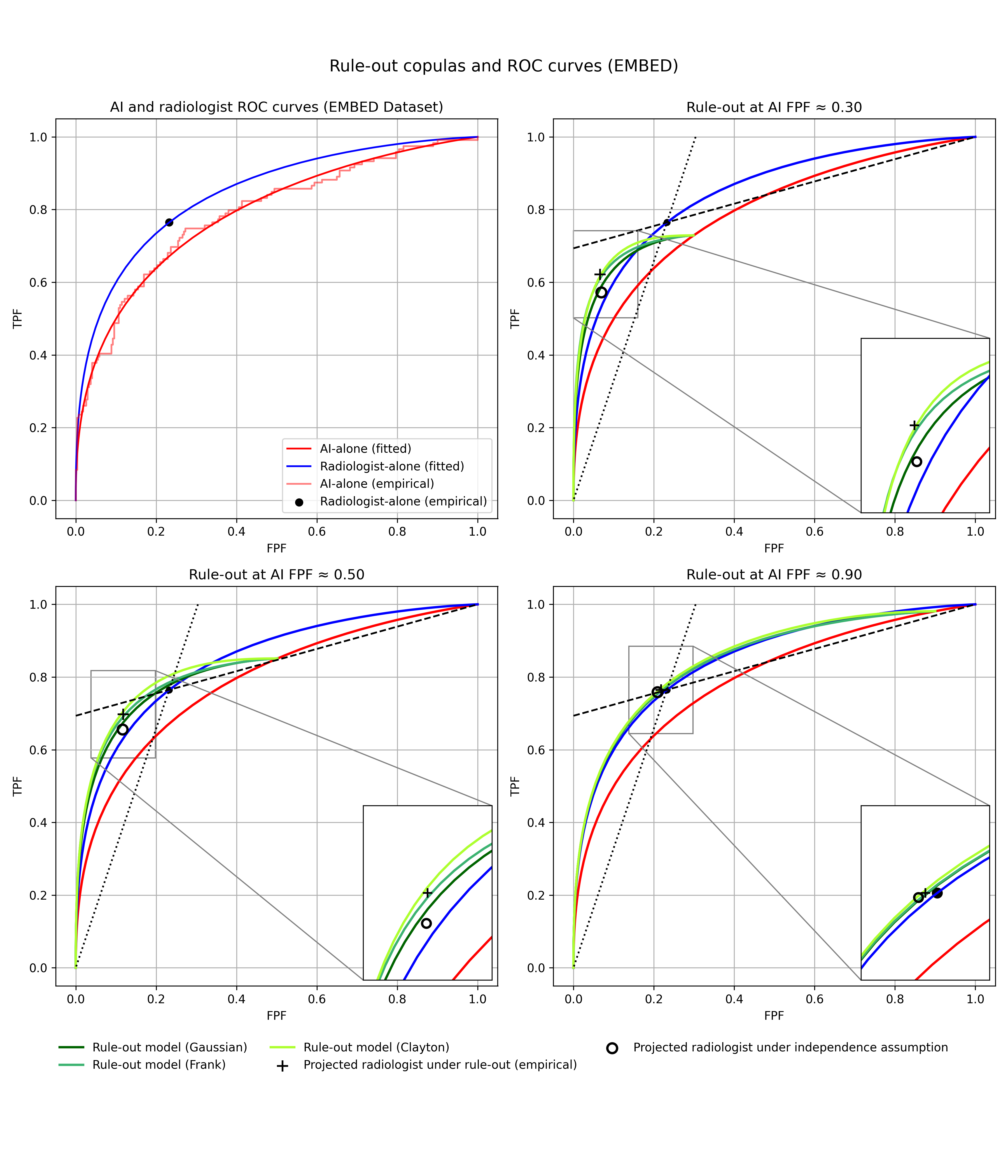}
\end{center}
\caption{The top-left plot gives both empirical and fitted radiologist- and AI-alone ROC curves on the EMBED testing dataset. The remaining plots include the radiologist-alone (Test B) and AI-alone (Test A) ROC curves on the EMBED testing set as well as models of rule-out ROC curves generated from Gaussian, Frank, and Clayton copulas, using the radiologist and AI marginals and score correlations as inputs. The three plots are generated with different AI rule-out operating thresholds at FPFs 0.3, 0.5 and 0.9 respectively. The empirical radiologist performance both with and without rule-out (at the same decision threshold) are included in each plot and in each case fall within the range of the proposed copula models. The lines of constant PPV and NPV equaling that of the radiologist are provided as dotted and dashed lines respectively for reference. For illustrative purposes, the projected radiologist performance is also plotted under an ``independence assumption," in which the radiologist and AI scores are assumed to have zero correlation, to show the less-optimal agreement between that point and the copula models. This independence assumption has been used in other works, e.g. \cite{Obuchowski}, but leads to an over-simplification of the models with worse agreement between simulation and theory. The independence model is the same regardless of copula.}
\label{fig:empirical_theory_compare_EMBED}
\end{figure}

The top-left plot in Figure \ref{fig:empirical_theory_compare_EMBED} shows the mean empirical radiologist performance, the empirical AI ROC curve, and the fitted ROC curves on the testing dataset. Given that we only have a single (FPF, TPF) point representing the average radiologist performance, we fit the full radiologist ROC curve using the assumption that the mean-to-sigma ratio~\cite{GreenNSwets} is the same as for the AI ROC curve.
Such an assumption is reasonable given that (1) both the AI and readers are processing the information in the same images using their respective neural networks and (2) the forms of published AI curves are similar to those of human readers.

In this example, 
the AI-alone performance is inferior to the radiologist-alone performance: the AUC for both the empirical and fitted AI ROC curves is 0.789, while the AUC for the fitted radiologist ROC curve is 0.847. Based on yes/no recall decisions, the radiologist (Se, Sp) is (0.763, 0.768). 

Pearson’s correlations between radiologist assessment and AI scores were 0.09 and 0.40 respectively for non-diseased and diseased images; Kendall’s tau correlations were 0.08 and 0.30. We note that these are rough measures due to the coarseness of the BIRADS scores provided by radiologists.

We set three hypothetical decision thresholds for the AI algorithm to be used as a rule-out test.  At each threshold we calculated the empirical radiologist rule-out operating points by calling all patients with AI scores below the threshold negative negative and then re-calculating the radiologist Se and Sp values; this re-calculation implicitly assumes that the radiologists would not change their decision thresholds in the presence of a rule-out AI and results in true and false positive rates that are always less than those of the radiologist alone.  
For these three thresholds the AI operated at (Se, Sp) of (0.723, 0.7), (0.85, 0.5) and (0.98, 0.1) respectively, and the empirical radiologist rule-out operating points (Se, Sp) were (0.622, 0.934), (0.697, 0.882) and (0.765, 0.782).

We then modeled a pair of bivariate distributions like Figure~\ref{fig:joint_copulas} that have the fitted parametric marginal distributions described above and copulas with the  empirically measured correlations.  We made models for the Gaussian, Frank, and Clayton copulas.  From these bivariate distributions we created ROC curves for rule-out and rule-in scenarios, and compared them to the empirical data.
Figure \ref{fig:empirical_theory_compare_EMBED} depicts the AI-alone and radiologist-alone ROC curves, the theoretical rule-out curves, and the empirical radiologist performance. 

Although the AI-alone ROC curve is uniformly lower than the radiologist-alone curve in this example, we see that coupling the AI with the radiologist in a rule-out scenario can lead to improved sensitivity for a range of FPF values. For example, in the top-right plot of Figure \ref{fig:empirical_theory_compare_EMBED}, the rule-out copula models are superior to the radiologist-alone model when the radiologist operates at an FPF between roughly 0 and 0.15.

The three copula model ROC curves (all in green in Figure \ref{fig:empirical_theory_compare_EMBED}) agree well with the average empirical rule-out performance (indicated by the + symbols in Figure \ref{fig:empirical_theory_compare_EMBED}). By comparison, the projected radiologist performance under an “independence assumption," in which the correlation between the radiologist and AI is assumed to be zero, agrees less well with the copula models. This illustrates the value of considering radiologist-AI correlation in models of joint rule-out performance.


In our calculations of the empirical radiologist performance under a rule-out scenario, we assumed that radiologists would operate at a fixed threshold with or without a prior rule-out AI. This is an unlikely assumption given that radiologists change their decision thresholds as disease prevalence changes \cite{GreenNSwets, Samuelson2013}. Thus, the estimated rule-out ROC curves are useful in providing a range of TPF and FPF values over which radiologists might operate if such a rule-out device were used in clinics.

\subsection{CSAW-CC Dataset}

The Cohort of Screen-Aged Women (CSAW) is a population-based cohort of women who were invited to mammography screening in the Stockholm region, SE, between 2008 and 2015 \cite{Dembrower}. Several datasets are part of the ecosystem of the CSAW cohort, including CSAW-M \cite{Sorkhei}, CSAW-CC \cite{CSAWCC}, and CSAW-S \cite{CSAWS}. The CSAW-CC dataset contains mammography x-ray images from breast cancer screening at the Karolinska University Hospital.
Because the CSAW-CC dataset includes individual radiologist recall decisions for two distinct radiologists on thousands of images, it is suitable for our analysis, despite the lack of a finer radiologist scoring scale.

\subsubsection{Dataset}

The publicly available dataset includes 873 cases of first-time breast cancer for women in the screening age range (40-74) and 7,850 healthy controls. 
Ground truth was established via biopsy for cancer cases. Non-cancer cases were reported as those who did not have a cancer diagnosis during the entire seven-year screening period. Individual radiologist recall decisions (yes/no) were reported by two distinct radiologists for each screening exam, as well as the joint consensus recall decision between the two radiologists.

We conducted the analysis on a per-patient-per-exam level. For the radiologist recall scores, we conditioned on one of the two radiologist decision columns to mimic a single-reader setting as in the United States. For each patient exam, we took the maximum AI score over all images read at that exam to be the final AI score. If any of the images read at that exam were found to be cancerous within 730 days, the patient was deemed cancer-positive at that exam.

\subsubsection{AI algorithm}

A commercial AI CAD system trained independently on external data was applied to the CSAW-CC dataset by F. Strand at Karolinska Institutet. No additional training was performed by the team at Karolinska.

\subsubsection{Analysis}

\begin{figure}[H]
\begin{center}
\includegraphics[trim=0cm 3cm 0cm 1cm, clip, width=14.5cm]{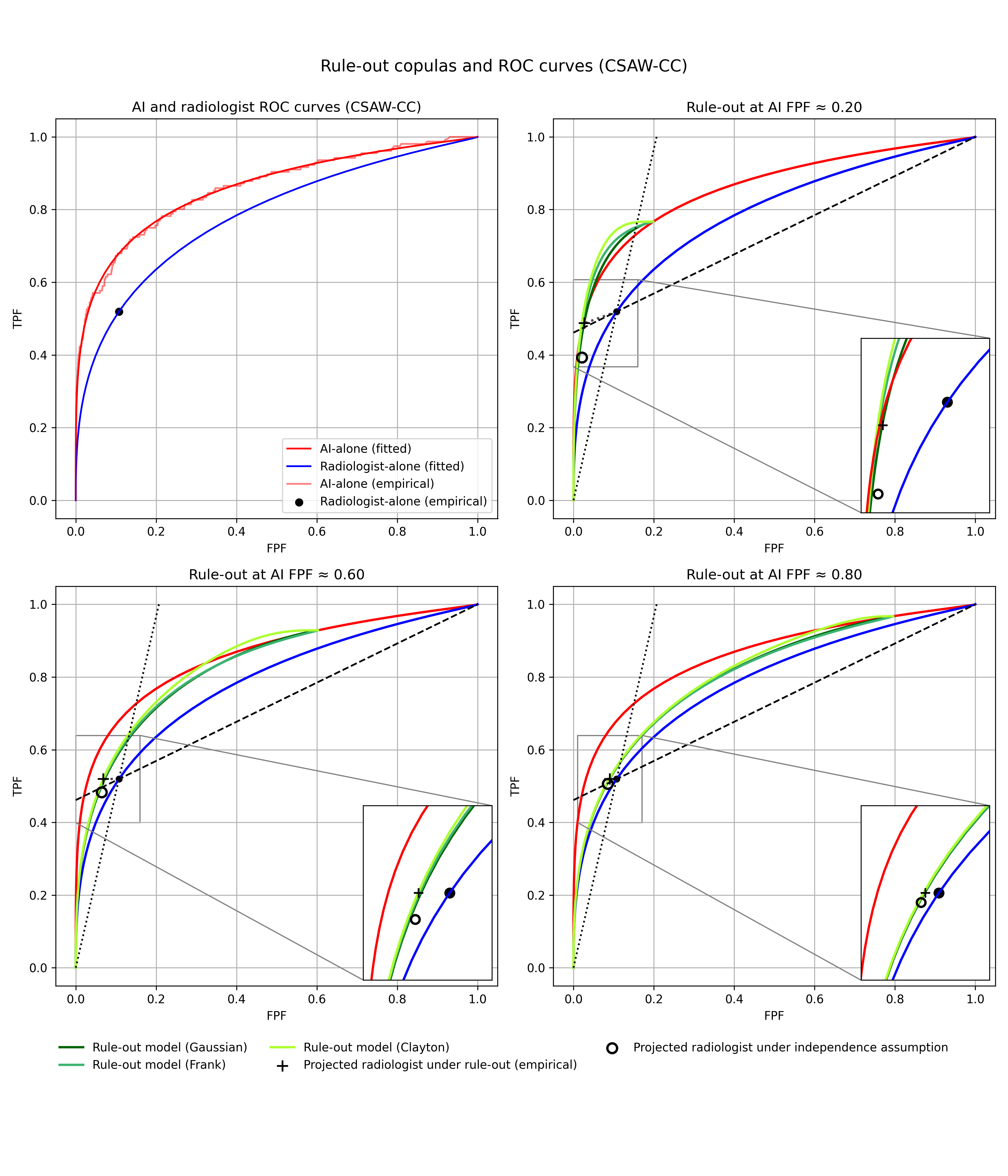}
\end{center}
\caption{The top-left plot gives both empirical and fitted radiologist- and AI-alone ROC curves on the CSAW-CC dataset. The remaining three plots include the fitted radiologist and AI curves for reference as well as models of rule-out-rule-in ROC curves generated from Gaussian, Frank, and Clayton copulas, using the radiologist and AI marginals and score correlations as inputs; the projected radiologist performance under a rule-out-rule-in scenario is also included. The three plots are generated with different AI operating thresholds at FPFs 0.2, 0.6 and 0.8 respectively. The empirical radiologist performance with and without rule-out (at the same decision threshold) are included in each plot and in each case fall within the range of the proposed copula models. The lines of constant PPV and NPV equaling that of the radiologist are provided as dotted and dashed lines respectively for reference. For illustrative purposes, the projected radiologist performance is again also plotted under the independence assumption, where the radiologist and AI scores are assumed to have zero correlation, to show less optimal agreement between that point and the copula models.}
\label{fig:empirical_theory_compare_CSAW}
\end{figure}

The top-left plot in Figure \ref{fig:empirical_theory_compare_CSAW} shows the average empirical radiologist performance and the empirical AI ROC curve, along with fitted binormal ROC curves for each. As in the previous subsection, we only have a single (FPF, TPF) point representing the radiologists' performance, so we fit the full radiologist ROC curve using the assumption that the mean-to-sigma ratio is the same as for the AI ROC curve.

Note that, in contrast to the EMBED example, the AI performance is superior to the radiologist performance on CSAW-CC: the AUC of both the fitted and empirical AI ROC curves is 0.91, and the AUC for the fitted radiologist curve is is 0.85. Based on yes/no recall decisions and with the inclusion of cancers found up to 730 days from screening, the radiologist FPR is 0.11 and the TPR is 0.66.

Pearson’s correlations between radiologist assessment and AI scores were 0.07 and 0.67 respectively for non-diseased and diseased images. Kendall’s tau correlations were 0.04 and 0.52. Note, again, that these are rough correlation measures due to the fact that the radiologist decision is binary.

Figure \ref{fig:empirical_theory_compare_CSAW}, like Figure \ref{fig:empirical_theory_compare_EMBED}, includes three plots of rule-out curves along with empirical radiologist performance at different AI operating points. In each case, the empirical performance values again fall within the range of the three copula models plotted. In this example, the empirical PPV and NPV of the radiologist under a rule-out scenario are superior to that for the PPV and NPV \emph{without} rule-out at each of the plotted AI operating points.
This contrasts with the EMBED example, where only for the highest AI FPF operating point had superior NPV and PPV due to the inferior performance of the AI compared to the radiologist in that case.

At the lowest AI FPF value of 0.2, the rule-out method on CSAW-CC achieves higher sensitivity than both the radiologist-alone and AI-alone diagnostics when operating at a fixed specificity over a range of FPF values (0.03 to 0.2). In contrast, at AI operating FPF values of 0.6 and 0.8 -- which represent more conservative scenarios with fewer cases automatically ruled-out by the AI diagnostic -- the rule-out curves generally fall short of the AI-alone performance. Nonetheless, because the AI significantly enhances the radiologist’s performance, the combined rule-out models outperform the radiologist in every scenario.

The CSAW-CC dataset also includes percent breast density scores as estimated by the Libra software version 1.0.4 \cite{Libra}. In Table \ref{table:CSAWdensity}, we provide several metrics on different density categories. The four Libra density categories (0-7.5, 7.5-22, 22-46, and 46-100) were chosen to roughly match the typical percentage of patients in each of the BIRADS density categories A, B, C and D respectively. Both the AI-alone and radiologist-alone AUCs decrease as density increases, as expected. The radiologist sensitivity and specificity are reported both without and with rule-out, where the AI operating threshold is set to an FPF of 0.5 on the full dataset. Specificity under rule-out increases substantially on each of the sets, with the highest increase on the lowest-density set (a difference of 0.071). Sensitivity decreases by at most 0.002, with one false positive in each of the middle two density categories and zero in the lowest and highest categories.

\begin{table}
\centering
    \includegraphics[scale=0.7]{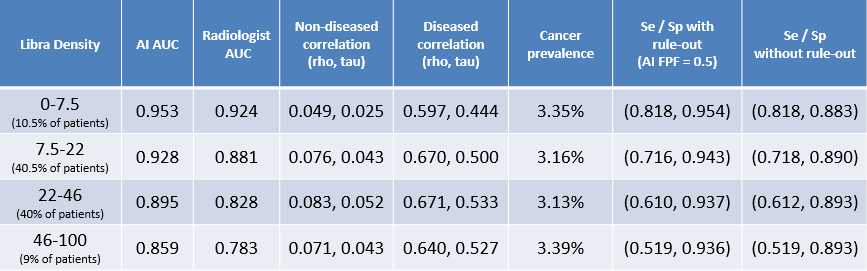}
    \caption{CSAW-CC metrics, stratified by density. }
    \label{table:CSAWdensity}
\end{table}

\subsection{Radboud University Reader Study}

The third dataset comes from a retrospective reader study with screening mammograms from the Dutch National Breast Cancer Screening Program conducted at Radboud University in which radiologist scores were collected indicating estimated probability of malignancy. Further details of the study are described below and in ~\cite{Gommers}.

\subsubsection{Reader Study Dataset}
A cancer-enriched dataset of anonymized two-view bilateral mammograms was compiled from 150 women, 75 of whom had a confirmed diagnosis of breast cancer. All mammograms were acquired between September 2016 and May 2019 using digital mammography systems from Hologic (Lorad Selenia and Selenia Dimensions). In cancer cases, ground truth was established via biopsy. For non-cancer cases, the absence of malignancy was confirmed through at least two years of negative follow-up.

The study included 12 screening-certified breast radiologists from different reading units in the Netherlands. The study was fully-crossed, so each reader assessed all 150 mammograms in the enriched dataset. Readers were asked to use a 0-to-100 slider to indicate the likelihood of malignancy, with 100 indicating the highest likelihood. In addition, readers recorded a binary decision (“recall” or “no recall”) for each case.

\subsubsection{AI algorithm}
The AI software used in the reader study was Transpara v.2.1.0. The algorithm divided each mammogram into four distinct regions and generated a score for each region. The overall case-level output was defined as the maximum score across these regions. The AI tool was trained on external data by the vendor, and no additional training was performed by the team at Radboud University.

\subsubsection{Rule-out and Rule-in ROC Analysis}

The advantage of the Radboud reader study dataset, compared to the EMBED and CSAW-CC datasets, is the finer scoring system, which allows us to obtain a more complete empirical ROC curve for the radiologist to which we may fit a binormal parametric curve. For the purposes of this analysis, we focus specifically on the performance of a single radiologist given potential drawbacks with averaging the radiologist performance~\cite{Chen2014}. To reflect scenarios with an average radiologist, we sampled the radiologist whose empirical AUC was close to the median: the 6th highest of the 12 radiologists. The AUC for the empirical radiologist ROC curve is 0.934 and for the fitted curve is 0.921. Pearson’s correlations between radiologist assessment and AI scores were 0.29 and 0.45 respectively for non-diseased and diseased images; Kendall’s tau correlations were 0.18 and 0.44. Figure \ref{fig:empirical_theory_compare_Radboud} depicts the theoretical rule-out-rule-in curves along with the empirical radiologist performance. 
\begin{figure}[H]
\begin{center}
\includegraphics[trim=0cm 3cm 0cm 1cm, clip, width=15cm]{rule_out_rule_in_radboud.png}
\end{center}
\caption{The top-left plot gives both empirical and fitted radiologist- and AI-alone ROC curves on the Radboud dataset. The radiologist is the  median-performing radiologist (6 out of 12) based on the empirical AUC value. The remaining three plots include the fitted radiologist and AI curves for reference as well as models of rule-out-rule-in ROC curves generated from Gaussian, Frank, and Clayton copulas, using the radiologist and AI marginals and score correlations as inputs; the projected radiologist performance under a rule-out-rule-in scenario is also included. The top-right plot models a rule-out-only scenario, with AI FPF 0.45. The bottom-left plot models a combined scenario in which the rule-out FPF threshold is set to 0.45 and the rule-in threshold is 0.05. The bottom-right plot models a rule-in-only scenario with FPF threshold at 0.05. The empirical radiologist ROC curves under the respective rule-out-rule-in scenarios are included in each plot as the dotted black curve. One can again see the strong agreement between the empirical performance and the copula models. The ``independence'' copula is plotted for illustrative purposes, showing again that use of true correlations leads to better agreement between the empirical and theoretical curves.}
\label{fig:empirical_theory_compare_Radboud}
\end{figure}

Figure \ref{fig:empirical_theory_compare_Radboud}, unlike Figures \ref{fig:empirical_theory_compare_EMBED} and \ref{fig:empirical_theory_compare_CSAW}, also includes the \emph{rule-in setting}, modeling a rule-out-only scenario in the top-right, a rule-out-rule-in scenario in the bottom-left, and a rule-in-only scenario in the bottom right. The projected empirical radiologist ROC curve under each scenario is plotted by automatically calling cases with AI score below the rule-out threshold negative and automatically calling cases above the rule-in threshold positive, then re-calculating the FPF and TPF at each threshold value. In each scenario, the empirical curve shows strong agreement with the three copula models plotted. Further, the ROC curves constructed assuming zero correlation (dashed gray line in Figure 6 with an independence assumption) again do not accurately model the empirical performance (dotted gray line in Figure 6).

In all scenarios modeled in Figure \ref{fig:empirical_theory_compare_Radboud}, the radiologist-and-AI performance is superior to that of the radiologist alone but still remains inferior to that of the AI alone given the substantial gap in performance between the AI and radiologist. 

Taken together with the EMBED and CSAW-CC examples, these findings indicate that rule-out-rule-in devices can augment radiologist performance even in settings where the AI does not outperform the radiologist. When the AI exhibits a large performance advantage, rule-out–rule-in strategies can yield substantial gains over radiologist-alone operation, even if no joint strategy exceeds the AI-alone diagnostic. When AI and radiologist performance are more closely matched, joint strategies can expand the achievable operating region beyond that of either diagnostic alone. 

\section{Discussion}
\label{sec:discussion}
In this work, we have introduced a theoretical methodology that uses bivariate copulas to construct and analyze ROC curves for combined rule-out (“believe the negative”) and rule-in (“believe the positive”) diagnostic strategies. Unlike prior methods that have assumed independence between AI and radiologist diagnostics, our method incorporates a correlation structure on both diseased and non-diseased cases.

A key insight from our analysis is that the partial area under the ROC curve (pAUC) in the rule-out scenario 
increases with higher radiologist-AI correlation among diseased cases and decreases with higher correlation among non-diseased cases. Conversely, the rule-in scenario exhibits the opposite behavior. These findings highlight the importance of the underlying correlation structure: when the radiologist and AI tend to agree on true positives (i.e., higher correlation in diseased cases), a rule-out strategy can improve sensitivity without sacrificing specificity. Conversely, in rule-in settings, lower diseased correlation and higher non-diseased correlation enhance diagnostic performance. Such insights may help determine optimal operating thresholds for AI devices and when designing integrated diagnostic workflows if a rule-out device were used in clinics.

One advantage of developing ROC theory is that ROC curves are very flexible. There is a somewhat understudied phenomenon that readers may change decision thresholds with the presence of an AI device. In particular, among the patients seen by the radiologist, a larger percentage may be recalled because those patients seen are known to the reader to be not a high-confidence negative. However, recall rate may decrease overall because a rule-out diagnostic automatically leads to fewer false positives. This is evident in a study by conducted over two years at three sites in Denmark~\cite{Lauritzen} in which an AI system was implemented midway through the study to replace one of two radiologists.

We validated our ROC theory with multiple mammography datasets, including EMBED, CSAW-CC, and a Radboud University reader study. In each case, the copula-based ROC models show strong agreement with empirical performance, and they outperform models based on the assumption of radiologist-AI independence. This highlights that accurately estimating the correlation between AI outputs and radiologist scores contributes to a more reliable rule-out and rule-in performance estimation.

Our copula approach provides a structured way to evaluate and compare different AI devices and radiologist-AI integration schemes if a rule-out device were used in clinics. The methodology can be applied to combinations of almost any two diagnostics, including rule-out/rule-in scenarios using two readers or readers with AI systems.  We expect that radiologists recall thresholds and rates would change upon the implementation of an AI rule-out or rule-in device.  The extended ROC analysis presented in this work provides a structured method to estimate performance across a range of radiologist recall rates.   


\section{Acknowledgments and Disclaimer}
\label{sec:disclaimer}
This work was supported by the FDA Office of Women’s Health. This project was supported in part by an appointment to the ORISE Research Participation Program at the Center for Devices and Radiological Health (CDRH), U.S. Food and Drug Administration, administered by the Oak Ridge Institute for Science and Education through an interagency agreement between the U.S. Department of Energy and FDA/Center.

This article reflects the views of the authors and does not represent the views or policy of the U.S. Food and Drug Administration, the Department of Health and Human Services, or the U.S. Government.  The mention of commercial products, their sources, or their use in connection with material reported herein is not to be construed as either an actual or implied endorsement of such products by the Department of Health and Human Services.

\appendix
\section{Calculus in copula analysis} 
\subsection{Gumbel copula: Equation \eqref{eq:TPFruleout_gumbel} is increasing in $\theta_{\D}$} 
\label{subsec:A1}
To see that \eqref{eq:TPFruleout_gumbel} is increasing in $\theta_{\D}$ on $(0, \infty)$, note first that $- \log F_{r\D}(x)$ and $-\log F_{a\D}(T_{\A})$ must be nonnegative as $F_{x \D}$ and $F_{y \D}$ are CDFs with range in $[0,1]$. Letting $- \log F_{r\D}(x) = a$ and $-\log F_{a\D}(T_{\A}) =b$ below,
\begin{align}
    \frac{d}{d\theta} &\exp(-(a^{\theta}+b^{\theta})^{1/\theta})  \\&=\exp(-(a^{\theta}+b^{\theta})^{1/\theta}) \cdot (a^{\theta} + b^{\theta})^{1/\theta} \cdot \Big(\frac{\log(a^{\theta}+b^{\theta})}{\theta^2} - \frac{a^{\theta}\log a + b^{\theta} \log b}{\theta(a^{\theta} + b^{\theta})}\Big).
\end{align}
This expression is positive when $\Big(\frac{\log(a^{\theta}+b^{\theta})}{\theta^2} - \frac{a^{\theta}\log a + b^{\theta} \log b}{\theta(a^{\theta} + b^{\theta})}\Big)$ is positive, which is the case if and only if
\begin{align}a^{\theta}\log(a^{\theta}+b^{\theta}) + b^{\theta}\log(a^{\theta}+b^{\theta}) &> a^{\theta}\cdot \theta \log a + b^{\theta} \cdot \theta \log b\\
&= a^{\theta} \log a^{\theta} + b^{\theta} \log b^{\theta}.
\end{align}
This is clearly true for all $\theta$, since $a, b > 0$ and thus $\log(a^{\theta}+b^{\theta}) > \max(\log a^{\theta}, \log b^{\theta}).$\\



\subsection{Frank copula: Kendall's tau is increasing in $\theta$}
\label{subsec:A2}
To see that Kendall's tau increases as $\theta$ increases for the Frank copula, for $0 < \theta < 2\pi$, we can use the series expansion for the Debye function to rewrite \eqref{eq:taufrank} as 
\begin{align}
    \tau_{\mathrm{Frank}}(\theta) = 4\sum_{k=1}^{\infty} \frac{B_{2k}}{(2k+1)!} \theta^{2k-1} = 1 + \frac{4}{\theta}\left(\frac{1}{\theta} \int_0^{\theta} \frac{t}{e^t-1} dt\right)-\frac{4}{\theta}.
\end{align}
Taking the derivative,
\begin{align} 
    \frac{d}{d\theta} \tau_{\mathrm{Frank}}(\theta) & = \label{eq:derivoftauFrank}
    \frac{d}{d\theta}\left(1 + \frac{4}{\theta}\left(\frac{1}{\theta} \int_0^{\theta} \frac{t}{e^t-1} dt\right)-\frac{4}{\theta}\right)\\
    & = \frac{4}{\theta^2} \frac{\theta}{e^{\theta}-1} + \frac{-8}{\theta^3}\int_0^{\theta} \frac{t}{e^t-1} dt + \frac{4}{\theta^2},
    \end{align}
which is positive if and only if 
\begin{align}
   F(\theta) := \theta + \frac{\theta^2}{e^{\theta}-1} - 2 \int_0^{\theta} \frac{t}{e^t-1} dt
\end{align}
is positive. Differentiating the above, we have
\begin{align}
    F'(\theta) & = 1 - \frac{\theta^2 e^{\theta}}{(1-e^{\theta})^2} = 1 - \Big(\frac{\theta}{2\sinh(\theta/2)}\Big)^2.
\end{align}
Using the general inequality $\sinh(t) > t$ for $t > 0$, we see that $\Big(\frac{\theta}{2\sinh(\theta/2)}\Big)^2 < 1$ for all $\theta > 0$. Thus, $F'(\theta) > 0$ for all $\theta > 0$ and $F$ is strictly increasing on $(0, \infty)$. Since $F(0) = 0$, we conclude $F(\theta) > 0$ for all $\theta > 0$ and therefore $\tau_{\mathrm{Frank}}(\theta)$ is positive for all $\theta > 0$.


\subsection{Frank copula: Equation \eqref{eq:TPFruleout_frank} is non-decreasing in $\theta$} 
\label{subsec:A3}
To see that \eqref{eq:TPFruleout_frank} is non-decreasing in $\theta$ for $\theta > 0$, note that
\begin{align}
& \frac{d}{d\theta} \Big(-\frac{1}{\theta}\log \Big[1+ \frac{(e^{-\theta u}-1)(e^{-\theta v}-1)}{e^{-\theta}-1}\Big]\Big) \nonumber\\ \label{eq:frankcopulaanalysis}
&= \log \Big(\frac{(e^{-u\theta}-1)(e^{-v\theta}-1)}{\theta^2}+1\Big) \\
&+ { \big(\frac{u e^{-u\theta}(e^{-v\theta}-1)}{e^{-\theta}-1}+\frac{ve^{-v\theta}(e^{-u\theta}-1)}{e^{-\theta}-1}-\frac{e^{-\theta}(e^{-u\theta}-1)(e^{-v\theta}-1)}{(e^{-\theta}-1)^2}\big)}{\theta^{-1}\left(\frac{(e^{-u\theta}-1)(e^{-v\theta}-1)}{e^{-\theta}-1}+1\right)^{-1}} \nonumber
\end{align}
The logarithm term above is always positive since $\frac{(e^{-u\theta}-1)(e^{-v\theta}-1)}{\theta^2}$ is positive for $u, v > 0$, and the denominator of the second term is positive for $u, v \in [0,1]$, which is the case in \eqref{eq:TPFruleout_frank} as $u$ and $v$ are CDFs. Thus, the second term is positive when
$$\frac{e^{-\theta}(e^{-u\theta}-1)(e^{-v\theta}-1)}{(e^{-\theta}-1)^2} \le \frac{u e^{-u\theta}(e^{-v\theta}-1)}{e^{-\theta}-1}+\frac{ve^{-v\theta}(e^{-u\theta}-1)}{e^{-\theta}-1}$$
or equivalently when
$$e^{-\theta}(e^{-u\theta}-1)(e^{-v\theta}-1) \le ue^{-u\theta}(e^{-v\theta}-1)(e^{-\theta}-1) + ve^{-v\theta}(e^{-u\theta}-1)(e^{-\theta}-1)$$
which is equivalent to the condition
$$1 < u\frac{e^\theta-1}{e^{u\theta}-1} + v\frac{e^\theta-1}{e^{v\theta}-1}.$$
We analyze the stronger condition that
$$1 \le u\frac{e^\theta-1}{e^{u\theta}-1}$$
This is equivalent to
$$e^{u\theta}-1  + (1-e^\theta) \le ue^\theta-u+(1-e^\theta)$$
or 
$$e^{u\theta}-e^\theta \le (e^\theta-1)(u-1).$$
Using the series expansion for $e^\theta$ at $\theta = 0$ (which converges everywhere), the above is equivalent to
$$\sum_{k=1}^{\infty} \frac{(u\theta)^k - \theta^k}{k!} \le \sum_{k=1}^{\infty}\frac{u\theta^k-\theta^k}{k!}$$
which is clearly true since $u \in [0,1]$.


\begin{thebibliography}{10}

\bibitem{Burns}
D.M. Burns, ``Primary prevention, smoking, and smoking cessation," \emph{Prevention and Early Diagnosis of Lung Cancer}, vol. 89, no. 11, 2000. 

\bibitem{Chen2014} W. Chen and F.W. Samuelson, ``The average receiver operating characteristic curve in multireader multicase imaging studies," \emph{British Journal of Radiology}, vol. 87, no. 1040, 2014. 

\bibitem{Dembrower} K. Dembrower, P. Lindholm and F. Strand, ``Multi-million Mammography Image Dataset and Population-Based Screening Cohort for the Training and Evaluation of Deep Neural Networks—the Cohort of Screen-Aged Women (CSAW)," \emph{Journal of Digital Imaging}, vol. 33, pp. 408-413, 2020. 

\bibitem{Dorfman}
D.D. Dorfman and E. Alf, ``Maximum-likelihood estimation of parameters of signal-detection theory and determination of confidence intervals—Rating-method data," \emph{Journal of Mathematical Psychology}, vol. 6, no. 3, pp. 487-496, 1969. 

\bibitem{DW}
Z. Dresner and G.O. Wesolowsky, ``On the computation of the bivariate normal integral," \emph{Journal of Statistical Computation and Simulation}, vol. 35, no. 1-2, pp. 101-107, 1990. 

\bibitem{EMBED_opensource}
EMBED Open Source Dataset, 2023. [Online]. Available: \url{https://github.com/Emory-HITI/EMBED_Open_Data}.

\bibitem{Esserman}
L.J. Esserman et al., ``The WISDOM Study: breaking the deadlock in the breast cancer screening debate," \emph{npj Breast Cancer}, vol. 3, p. 34, 2017. 

\bibitem{GM}
C. Genest and J. MacKay, ``The joy of copulas: Bivariate distributions with uniform marginals," \emph{The American Statistician}, vol. 40, no. 4, pp. 280-283, 1986. 

\bibitem{GF}
A. Genz and F. Bretz, ``Computation of Multivariate Normal and t Probabilities,'' \emph{Springer Science and Business Media}, 2009. 

\bibitem{Gommers} J.J.J. Gommers et al., ``Enhancing Radiologist Reading Performance by Ordering Screening Mammograms Based on Characteristics That Promote Visual Adaptation," \emph{Radiology}, vol. 313, no. 1, 2024. 


\bibitem{GreenNSwets} D. M. Green and J. A. Swets, ``Signal Detection Theory and Psychophysics," John Wiley \& Sons, New York, 1966.

\bibitem{HM}
J.A. Hanley and B.J. McNeil, ``The meaning and use of the area under a Receiver Operating Characteristic (ROC) curve," \emph{Radiology}, vol. 143, no. 1, pp. 29-36, 1982. 

\bibitem{Hickman}
S.E. Hickman et al., ``Mammography Breast Cancer Screening Triage Using Deep Learning: A UK Retrospective Study," \emph{Radiology}, vol. 309, no. 2, 2023.

\bibitem{Hofert}
M. Hofert, M. Maechler and A. J. McNeil, ``Likelihood inference for Archimedean copulas in high dimensions under known margins," \emph{Journal of Multivariate Analysis}, vol. 110, pp. 133-150, 2012. 

\bibitem{EMBED}
J. Jeong et al., "The EMory BrEast imaging Dataset (EMBED): A racially diverse, granular dataset of 3.4 million screening and diagnostic mammographic images," Radiology: Artificial Intelligence, vol. 4, no. 5, 2023.

\bibitem{Khreich}
W. Khreich, ``Towards adaptive anomaly detection systems using Boolean combination of hidden Markov models'', 2011. (Doctoral dissertation, École de technologie supérieure, Université du Québec, Montréal, Canada).

\bibitem{Kyono}
T. Kyono, F. Gilbert and M. van der Schaar, "Improving workflow efficiency for mammography using machine learning," Journal of the American College of radiology, vol. 17, pp. 56-63, 2020. 


\bibitem{Lang}
K. Lang, ``Artificial intelligence-supported screen reading versus standard double reading in the Mammography Screening with Artificial Intelligence trial (MASAI): a clinical safety analysis of a randomised, controlled, non-inferiority, single-blinded, screening accuracy" \emph{The Lancet Digital Health}, vol. 5, no. 7, pp. e400-e409, 2023. 


\bibitem{Larsen}
M. Larsen, ``Artificial intelligence evaluation of 122,969 mammography examinations from a population- based screening program," \emph{Radiology}, vol. 303, pp. 619-626, 2022. 

\bibitem{Lauritzen} A.D. Lauritzen et al., ``An `Artificial Intelligence–based Mammography Screening Protocol for Breast Cancer: Outcome and Radiologist Workload," \emph{Radiology}, vol. 304, no. 1, 2022. 

\bibitem{Libra} Libra Software, 1.0.4, Philadelphia, PA: University of Pennsylvania. 

\bibitem{Louro}
J. Louro, M. Posso, M. H. Boon, M. Román, L. Domingo, and M. Sala, ``A systematic review and quality assessment of individualised breast cancer risk prediction models," \emph{British Journal of Cancer}, vol. 121, no. 1, pp. 76–85, 2019.

\bibitem{Mahkonen}
M. Mähkönen, T. Virtanen, and J. Kämäräinen, ``Cascade of Boolean detector combinations,'' \emph{EURASIP Journal on Image and Video Processing}, vol. 2018, issue 61, 2018.

\bibitem{Marshall}
R. J. Marshall, ``The analysis of case-control data with clustered exposures," \emph{Biometrics}, vol. 57, no. 3, pp. 712–719, 2001.

\bibitem{CSAWS} C. Matsoukas, ``CSAW-S: mammography screenings from 172 different patients with annotations for semantic segmentation," 2023. [Online]. Available: \url{https://github.com/ChrisMats/CSAW-S}.


\bibitem{MazoGirard}
G. Mazo and S. Girard, ``A flexible and tractable class of one-factor copulas," \emph{Statistics and Computing}, vol. 1, pp. 1-16, 2015. 

\bibitem{McClish}
D.K. McClish, A. Wilk and C. Schubert, ``Choosing between the BP and BN sequential strategies. Pharmaceutical Statistics," Pharmaceutical Statistics, vol. 18, no. 5, pp. 533-545, 2019. 

\bibitem{McKinney}
S.M. McKinney, ``International evaluation of an AI system for breast cancer screening," \emph{Nature}, vol. 577, no. 7788, pp. 89-94, 2020. 

\bibitem{Metz}
C.E. Metz, ``Basic principles of ROC analysis," \emph{Seminars in Nuclear Medicine}, vol. 8, no. 4, pp. 283-298, 1978. 

\bibitem{Metz2006}
C.E. Metz, ``ROC analysis in medical imaging: a tutorial review of the literature," \emph{Radiological Physics and Technology}, vol. 1, pp. 2-12, 2008. 

\bibitem{Metz2007} C.E. Metz, ``Receiver operating characteristic analysis: a tool for the quantitative evaluation of observer performance and imaging systems," \emph{Journal of the American College of Radiology}, vol. 3, no. 6, pp. 413-422, 2006. 

\bibitem{CORROC} C.E. Metz, P.L. Wang and H.B. Kronman, ``A new approach for testing the significance of differences between ROC curves from correlated data," in \emph{Information Processing in Medical Imaging}, Dordrecht, Springer, 1984, pp. 432-445.ue; 1984, pp. 432–445. 

\bibitem{Naifar}
N. Naifar, ``Modeling dependence structure with Archimedean copulas and applications to the iTraxx CDS index," \emph{Review of Quantitative Finance and Accounting}, vol. 38, no. 2, pp. 271–294, 2012.

\bibitem{Nelsen} R. Nelsen, An Introduction to Copulas (Second Edition), Portland, OR: Springer Series in Statistics, 2006. 

\bibitem{Obuchowski}
N. A. Obuchowski and J. A. Bullen, ``Statistical considerations for testing an AI algorithm used for prescreening lung CT images," \emph{Translational Lung Cancer Research}, vol. 8, no. 2, pp. 179-193, 2019. 

\bibitem{Qin}
Z.Z. Qin et al., ``Using artificial intelligence to read chest radiographs for tuberculosis detection: A multi-site evaluation of the diagnostic accuracy of three deep learning systems," \emph{Scientific Reports}, vol. 9, 2019.

\bibitem{RSNAKaggle}
"RSNA Breast Cancer Detection Competition," 2023. [Online]. Available: \url{https://www.kaggle.com/competitions/ rsna-breast-cancer-detection}.

\bibitem{RayaPovedano}
J.L. Raya-Povedano, S. Romero-Martín, E. Elías-Cabot, A. Gubern-Mérida, M. Rodríguez-Ruiz and M. Álvarez-Benito, ``AI-based strategies to reduce workload in breast cancer screening with mammography and tomosynthesis: A retrospective evaluation," \emph{Radiology}, vol. 299, no. 1, pp. 50-57, 2021. 

\bibitem{RodriguezRuiz}
A. Rodríguez-Ruiz et al., ``Can we reduce the workload of mammographic screening by automatic identification of normal exams with artificial intelligence? A feasibility study," European \emph{Radiology}, vol. 29, no. 9, pp. 4825-4832, 2019. 

\bibitem{RodriguezRuiz2}
A. Rodríguez-Ruiz et al., ``Stand-alone artificial intelligence for breast cancer detection in mammography: Comparison with 101 radiologists," \emph{JAMA Network Open}, vol. 2, no. 3, 2019. 

\bibitem{Samuelson2013}
F.W. Samuelson, ``Inference Based on Diagnostic Measures from Studies of New Imaging Devices'',  \emph{Academic Radiology}, vol. 20, pp. 816-824, 2013. 

\bibitem{Santeramo} R. Santeramo, C. Damiani, J. Wei, G. Montana and A. R. Brentnall, ``Are better AI algorithms for breast cancer detection also better at predicting risk? A paired case–control study," \emph{Breast Cancer Research}, vol. 26, no. 25, 2024. 

\bibitem{Shen}
C. Shen, ``On the principles of believe the positive and believe the negative for diagnosis using two continuous tests," \emph{Journal of Data Science}, vol. 6, pp. 189-205, 2008. 

\bibitem{Sorkhei}
M. Sorkhei et al., ``CSAW-M: An Ordinal Classification Dataset for Benchmarking Mammographic Masking of Cancer," in \emph{NeurIPS 2025 Proceedings}.

\bibitem{CSAWCC} F. Strand, ``CSAW-CC (mammography) – a dataset for AI research to improve screening, diagnostics and prognostics of breast cancer (Version 1) [Dataset]," Solna, Sweden, 2022.

\bibitem{Takeuchi}
T. Takeuchi, ``Constructing a bivariate distribution function with given marginals and correlation: application to the galaxy luminosity function," \emph{Monthly Notices of the Royal Astronomical Society}, vol. 406, no. 3, 2010. 

\bibitem{Terry}
M. B. Terry et al., ``10-year performance of four models of breast cancer risk: A validation study," \emph{JAMA Oncology}, vol. 5, no. 11, pp. 1718-1723, 2019. 

\bibitem{Thompson}
M. L. Thompson, ``Assessing the diagnostic accuracy of a sequence of tests," \emph{Biostatistics}, vol. 4, no. 3, pp. 341-351, 2003. 

\bibitem{GV}
G.G. Venter, ``Tails of copulas," \emph{ASTIN Bulletin}, vol. 32, no. 2, pp. 259–270, 2002.

\bibitem{Yala}
A. Yala, C. Lehman, T. Schuster, T. Portnoi, and R. Barzilay, ``A deep learning mammography-based model for improved breast cancer risk prediction," \emph{Radiology}, vol. 292, no. 1, pp. 60–66, 2019.

\bibitem{HyunYouk}
J.H. Youk and E.K. Kim, ``Research highlight: Artificial intelligence for ruling out negative examinations in screening breast MRI," \emph{Korean Journal of Radiology}, vol. 23, no. 2, pp. 153-155, 2022. 

\bibitem{vanWinkel}
S.L. van Winkel et al., ``Impact of artificial intelligence support on accuracy and reading time in breast tomosynthesis image interpretation: A multi-reader multi-case study," \emph{European Radiology}, vol. 31, pp. 7348-7356, 2021.


\end{thebibliography}
\end{document}